# FreSCo: Mining Frequent Patterns in Simplicial Complexes

Giulia Preti
ISI Foundation, Italy
Turin, Italy
giulia.preti@isi.it

Gianmarco De Francisci Morales
ISI Foundation
Turin, Italy
gdfm@acm.org

Francesco Bonchi
ISI Foundation, Italy
Eurecat, Spain
francesco.bonchi@isi.it

## ABSTRACT

Simplicial complexes are a generalization of graphs that model higher-order relations. In this paper, we introduce simplicial patterns —that we call *simplets*— and generalize the task of frequent pattern mining from the realm of graphs to that of simplicial complexes. Our task is particularly challenging due to the enormous search space and the need for higher-order isomorphism. We show that finding the occurrences of simplets in a complex can be reduced to a bipartite graph isomorphism problem, in linear time and at most quadratic space. We then propose an anti-monotonic frequency measure that allows us to start the exploration from small simplets and stop expanding a simplet as soon as its frequency falls below the minimum frequency threshold. Equipped with these ideas and a clever data structure, we develop a memory-conscious algorithm that, by carefully exploiting the relationships among the simplices in the complex and among the simplets, achieves efficiency and scalability for our complex mining task. Our algorithm, FreSCo, comes in two flavors: it can compute the exact frequency of the simplets or, more quickly, it can determine whether a simplet is frequent, without having to compute the exact frequency. Experimental results prove the ability of FreSCo to mine frequent simplets in complexes of various size and dimension, and the significance of the simplets with respect to the traditional graph patterns.

## CCS CONCEPTS

• **Mathematics of computing** → **Hypergraphs**; *Graph algorithms*; • **Information systems** → *Data mining*.

## KEYWORDS

Frequent pattern mining, simplicial complex, higher order patterns.

**ACM Reference Format:**
Giulia Preti, Gianmarco De Francisci Morales, and Francesco Bonchi. FreSCo: Mining Frequent Patterns in Simplicial Complexes. https://doi.org/10.1145/3485447.3512191

## 1 INTRODUCTION

Frequent pattern mining in graph-structured data aims at finding structures that occur frequently in a given (set of) graph, under the assumption that frequency indicates relevance. This fundamental primitive has gained considerable attention thanks to the numerous applications such as fraud detection [37], network intrusion detection [22], trend discovery [44], Web usage mining [16, 23], link prediction for people-recommender systems, community discovery, or predicting group activity on social networks [11, 13, 34]. As the search space of the mining task is intrinsically exponential, the measure of frequency must be selected carefully. Typically, *anti-monotone measures* —such that a pattern cannot be more frequent that any of its sub-patterns [14, 20]— are the choice, as they allow to exploit the so-called *apriori* property to prune the search space substantially, thus enabling efficient algorithms.

In this paper, we generalize the task of frequent pattern mining from graphs —the case in which entities are involved in pairwise interactions— to the case of higher-order relations. In fact, many interactions in the real-world occur among more than two entities at once [15, 50]. For example, multiple Web pages can be visited in the same session, multiple entities co-occur in the same Wikipedia page, scientific papers are usually written by teams of researchers, and molecules interact in groups. Simple graphs are not expressive enough to model such higher-order relations, because they cannot distinguish the case of, e.g., three entities co-occurring in pairs in three different Wikipedia pages, from the case where all three entities co-occur in the same page [10].

*Hypergraphs* [17] and *simplicial complexes* [7] are higher-order generalizations of simple graphs that allow such a distinction [48]. Hypergraphs generalize graphs by allowing *hyperedges* to connect more than two vertices. Conversely, a simplicial complex is a collection of polytopes such as triangles and tetrahedra, which are called *simplices*. Both these structures can be used to represent any higher-order relation [47] and their distinction is thin: simplicial complexes require the *downward closure* property, i.e., every substructure (also known as *face*) of a simplex contained in a complex $\mathcal{K}$ is also in $\mathcal{K}$. Using the example of three entities co-occurring in a Wikipedia page, the downward closure property models the fact that any pair of the three entities are also co-occurring in the page. Which paradigm is more appropriate depends on the application domain and the semantics we want to associate to a higher-order relation. In domains characterized by interactions that are "maximal" [51], e.g., in scientific collaborations (the authors of a paper are all co-authors) or gene activation pathways (largest group of collectively activated genes), simplicial complexes are more appropriate. Indeed, simplicial complexes have proven to be useful in analyzing the organization of the brain [31], characterizing the scientific collaborations [39], predicting new links between entities [9], and studying the mechanisms of social contagion [26].

**Challenges and contributions.** Given a simplicial complex, we define the problem of extracting all frequent simplicial patterns where a simplicial pattern is a subcomplex that we dub *simplet* (in analogy to the *graphlet*). The frequency measure we adopt is inspired by the MNI support measure [20] to ensure anti-monotonicity and efficiency in the computation. Interestingly, when the input complex encodes only pairwise interactions (i.e., when it is a graph), our problem formulation reduces to the traditional frequent graph pattern mining. Frequent simplet mining can find applications in complex summarization and classification, in characterizing the interactions in brains of patients suffering different diseases, in finding relevant functional dependencies in ecological systems, in determining the coactivation patterns in resting-state functional magnetic resonance imaging signals [30], in developing recommending systems [25], and in community detection [12].

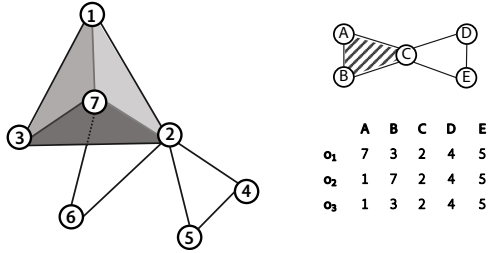

Figure 1: A simplicial complex (left), and a simplet with its occurrences (right).

Frequent pattern mining in simplicial complexes presents numerous challenges. First, the search space is extremely large, due to all the possible combinations of higher-order simplices that can constitute the simplet. Second, finding the occurrences of simplets in a complex entails determining if two simplets are *isomorphic*. While there exist many libraries for solving the graph isomorphism problem, little effort has been devoted to complex isomorphism [5]. We show that finding the occurrences of simplets in a complex can be reduced to a bipartite graph isomorphism problem, in linear time and at most quadratic space. We then propose an anti-monotonic frequency measure that allows us to start the exploration from small simplets and stop expanding a simplet as soon as its frequency falls below a minimum frequency threshold. Based on these ideas and a clever data structure, we devise a memory-conscious algorithm called FreSCo, that carefully exploits the relations among the simplices in the complex and among the simplets, to achieve efficiency. FreSCo comes in two flavors: it can compute the exact frequency of the simplets or, more quickly, it can determine whether a simplet is frequent, without computing the exact frequency.

Experiments on several real-world complexes prove the scalability of FreSCo in mining frequent simplets from complexes of various size and dimension. We also show the significance of the simplets with respect to the traditional graph patterns. In particular, we show that frequent simplets allow a finer-grained characterization of the dynamics happening among the entities in the complex. Finally, we showcase the usage of the extracted frequent simplets to build a method for *simplicial closure prediction* [9].

## 2 RELATED WORK

**Frequent pattern mining.** In the graph context, the problem of mining frequent or significant patterns has been widely studied in the literature. Existing works tackle the problem in unweighted graphs [1, 18, 27, 42], weighted graphs [43], streaming [6, 36], and uncertain graphs [49], among others. These works, however, cannot be adapted to the complex case straightforwardly, because they are not designed to **(i)** search patterns in our search space, and **(ii)** compute their frequency correctly. This is because graphs are not powerful enough to express higher-order structures.

Frequent itemset mining entails discovering all the subsets of items appearing frequently in a transactional database [2, 28, 35]. Given that a simplex can be represented as a set of vertices, a complex can be seen as a family of sets, i.e., a transactional database. Hence, the task of finding frequent simplets in a complex might resemble that of frequent itemset mining. However, the latter counts the number of transactions in the database that contain all the elements in the itemset, while in our setting, simplets are connected sub-complexes, and as such, they consist of many interconnected simplices that can span across multiple simplices in the complex.

**Frequent hypergraph mining.** Frequent pattern mining has been studied also in the context of hypergraphs, which are generalizations of graphs related to simplicial complexes. We recall that a complex is a special kind of hypergraph that satisfies the downward closure property. The seminal paper on frequent sub-hypergraph mining [25] proves that the problem is NP-hard, but can be solved in incremental polynomial time for node and edge injective hypergraphs. This work, however, finds frequent sub-hypergraphs in hypergraph databases, and hence cannot mine frequent structures in single hypergraphs, nor in single complexes. The reason is that, in the database setting, the frequency of a sub-hypergraph is defined as the number of hypergraphs in the database that contain a given sub-hypergraph, and hence, in a single hypergraph each sub-hypergraph would have frequency equal to either 1 or 0.

**Simplicial complexes analysis.** Simplicial complexes have been adopted to model higher-order relations and solve several interesting problems [8, 45]. By observing that the interactions between subsets of a group of users in a social network increase the likelihood that the members of the group will be pairwise connected in the future, Benson et al. [9] tackled link prediction via simplicial closure. Similarly, Eswaran et al. [19] addressed the semi-supervised learning task of label propagation in partially-labeled graphs. Iacopini et al. [26] adopted simplicial complexes to describe social contagion and diffusion phenomena, Horak et al. [24] to characterize networks by means of their topological features, Serrano and Gómez [46] to define centrality measures, Preti et al. [41] addressed the problem of truss decomposition of simplicial complexes. Finally, substantial work has been done by using simplicial complexes to study the brain's functional and structural organization [31, 40].

## 3 PROBLEM DEFINITION

We next introduce the concepts and the notation needed for our problem statement. After each definition, we use an example referring to Figure 1 to clarify the introduced concept. Let $V$ be a set of vertices. Simplices are higher-order generalizations of edges.

DEFINITION 1 (q-SIMPLEX, FACE, JOIST). *A q-dimensional simplex (q-simplex)* $\sigma^{(q)}$ *is a set of $q + 1$ interconnected elements of $V$, i.e., $\sigma^{(q)} = [v_0, \ldots, v_q] \subset V$. A face of $\sigma^{(q)}$ of dimension $r$ is an $r$-simplex $\sigma^{(r)} \subset \sigma^{(q)}$. The joist of $\sigma^{(q)}$ is the set $J_{\sigma^{(q)}}$ of all its $q + 1$ cofaces (faces of dimension $q - 1$) [41].*

EXAMPLE 1. *The dark triangle [2, 3, 7] in Figure 1 (left) is a 2-simplex, all its proper subsets of dimension 0 or 1 are its faces, while*

the set of 1-simplices $\{[2, 3], [2, 7], [3, 7]\}$ constitutes its joist. A 0-simplex is a relation involving a single vertex, and hence is equivalent to a vertex in a graph, while a 1-simplex is comparable to an edge, as they both relate two vertices. In contrast, a q-simplex with $q > 1$ is a polyadic relation that cannot be expressed in graph terminology.

DEFINITION 2 (COMPLEX). *A* simplicial complex $\mathcal{K}$ *is a set of simplices such that all the faces of the simplices in* $\mathcal{K}$ *are also in* $\mathcal{K}$. *The* dimension of $\mathcal{K}$ *is the dimension of its largest simplex, while its* vertex set $V_{\mathcal{K}}$ *is the set of distinct vertices in its simplices. The* n-skeleton *of* $\mathcal{K}$ *is the complex composed by the subset of simplices of dimensions $q \leq n$.*

EXAMPLE 2. *The complex in Figure 1 (left) has dimension 3, given by its 3-simplex $[1, 2, 3, 7]$. The 1-skeleton of a complex $\mathcal{K}$ corresponds to the undirected simple graph induced by the 1-simplices of $\mathcal{K}$. When the dimension of $\mathcal{K}$ is 1, the complex is a graph.*

A *subcomplex* $S_{\mathcal{K}} \subset \mathcal{K}$ is a downward-closed improper subset of simplices in $\mathcal{K}$ (the skeleton is also a subcomplex). We say that two simplices are *connected* if they share a face, and we call a connected subcomplex a *simplet*.

DEFINITION 3 (SIMPLET). *A simplet $P$ is a tuple $(V_P, C_P)$, where $C_P$ is a set of simplices such that **(i)** $\sigma^{(q)} \in C_P$ only if all its faces are in $C_P$, and **(ii)** for each pair of vertices $u, v$ in $V_P = \bigcup_{\sigma \in C_P} \{v \mid v \in \sigma\}$ there exists a sequence $[[u, u_0], [u_0, u_1], \ldots, [u_n, v]]$ of 1-simplices that connects $u$ and $v$. The* dimension *of the simplet $\dim(P)$ is the dimension of the largest simplex in $C_P$.*

EXAMPLE 3. *The structure in Figure 1 (right) is a simplet with vertex set $\{A, B, C, D, E\}$ and simplex set $\{[A, B, C], [C, D], [C, E], [D, E]\}$.*

DEFINITION 4 (ISOMORPHISM AND AUTOMORPHISM). *We say that a complex $\mathcal{K}_1$ is* isomorphic *to another complex $\mathcal{K}_2$, denoted as $\mathcal{K}_1 \simeq \mathcal{K}_2$, iff there exists a bijection $\phi : V_{\mathcal{K}_1} \mapsto V_{\mathcal{K}_2}$ that preserves the relations between the simplices, i.e., for each $\sigma = [v_0, \ldots, v_q] \in \mathcal{K}_1$ it holds that $\sigma' = [\phi(v_0), \ldots, \phi(v_q)] \in \mathcal{K}_2$. A bijection from a complex $\mathcal{K}$ to itself is called* automorphism, *and the set of all the automorphisms of $\mathcal{K}$ is denoted as $Auto(\mathcal{K})$. When the dimension of $\mathcal{K}$ is 1, these definitions are consistent with graph isomorphism and automorphism.*

DEFINITION 5 (OCCURRENCE). *The subcomplexes of $\mathcal{K}$ isomorphic to $P$ are called* occurrences *of $P$ in $\mathcal{K}$ (a.k.a. instances or embeddings).*

EXAMPLE 4. *The table in Figure 1 reports the occurrences of the simplet above the table, with the corresponding complex-vertex-to-simplet-vertex assignments. This simplet is characterized by one open and one closed triangle connected by a vertex. For example, the subcomplex $o_3 = (\{1, 2, 3, 4, 5\}, \{[1, 2, 3], [2, 4], [2, 5], [4, 5]\})$ is mapped to the simplet by the function $\phi(1) = A$, $\phi(2) = C$, $\phi(3) = B$, $\phi(4) = D$, $\phi(5) = E$, while no subcomplex including the edges of the triangle $[2, 6, 7]$ can be isomorphic to the simplet, because $[2, 6, 7]$ is an empty triangle that share an edge with any closed triangle in the complex.*

We are interested in finding simplets that occur frequently in the complex, with the assumption that frequency captures an important characteristic of the data-generating process. However, finding the occurrences of a simplet in a complex is a challenging task that requires the evaluation of isomorphisms between candidate subcomplexes and the simplet. Simplicial complexes are special kinds of hypergraphs, and the *SUB-HYPERGRAPH ISOMORPHISM* problem (SHI) is NP-hard by reduction from *SUBGRAPH ISOMORPHISM* (SGI) [32]. Given the hardness of the task, it is pivotal to design algorithms able to significantly prune the search space. A pruning strategy extensively used in graph pattern mining is based on the *anti-monotonicity* property, which states that the frequency of a super-pattern is smaller or equal to the frequency of its sub-patterns. Thanks to this property, the mining algorithm can follow a pattern-growth approach that starts from small structures and expands only the frequent ones. However, defining the support as the number of occurrences does not ensure anti-monotonicity [14]. Therefore, alternative support measures have been introduced in the literature, with *MNI* (Minimum Node-based Image) being the most used one, as it is more efficient to compute than the others [20].

To ensure anti-monotonicity and backward compatibility, we define a support measure *SUP* that is a proper generalization of the *MNI* measure to simplicial complexes:

DEFINITION 6 (SUP). *Let the* image set $I_P(v)$ *of* $v \in V_P$ *of a simplet $P$ be the set of vertices in $\mathcal{K}$ that are mapped to $v$ by some isomorphism $\phi$, i.e., $I_P(v) = \{u \in \mathcal{K} \mid \exists \phi \text{ s.t. } \phi(u) = v\}$. Then, $SUP(P, \mathcal{K}) = \min_{v \in V_P} |I_P(v)|$.*

PROPERTY 1. *SUP is anti-monotone.*

The proof of anti-monotonicity of *SUP* is in Appendix A.2.

We say that a simplet is *frequent* in $\mathcal{K}$ if $SUP(P, \mathcal{K}) \geq \tau$, where $\tau$ is a user-given minimum frequency threshold. Then, the problem addressed in this paper, i.e., *frequent simplet mining in simplicial complexes*, can be formalized as follows:

PROBLEM 1 (FREQUENT SIMPLET MINING). *Given a simplicial complex $\mathcal{K}$, a maximum size $s^*$, a minimum dimension $d^*$, and minimum frequency threshold $\tau$, find all the simplets $P$ such that $|V_P| \leq s^*$, $\dim(P) \geq d^*$, and $SUP(P, \mathcal{K}) \geq \tau$.*

Problem 1 is a decision problem: determining which simplets are frequent in a complex. The candidate simplets are all the possible simplets up to a given size and with minimum dimension. The minimum dimension leads to more compact but informative results, allowing the user to focus only on the most interesting higher-order structures in the complex. The maximum size limits the search space and hence controls the complexity of the mining task. This design choice is usually adopted also in frequent graph pattern mining [27], given the hardness of the task.

OBSERVATION 1. *Problem 1 is a proper generalization of traditional graph pattern mining. That is, when the simplicial complex $\mathcal{K}$ is a graph, Problem 1 with $d^* = 1$ corresponds exactly to graph pattern mining. In fact, a 1-dimensional simplet can be mapped, without loss, to a graph pattern, by mapping each 1-simplex to an edge and each 0-simplex to a vertex.*

## 4 GENERATION AND CANONICALITY

Before delving into our algorithm, we give an overview of how it generates the simplets to examine, and how it avoids examining the same simplet twice.

The simplet search space is a lattice, partially ordered by *complex inclusion*, denoted as $\leq_C$, and defined as follows:

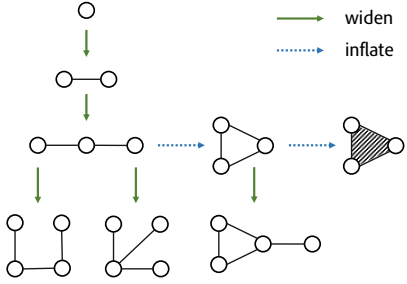

**Figure 2: Expansion of a simplet via application of widen and inflate. The first rule moves downward (size increase), while the second one moves rightward (dimension increase).**

DEFINITION 7 (COMPLEX INCLUSION). *Given two distinct simplets $P_1$ and $P_2$, we say that $P_1 \leq_C P_2$ iff for each simplex $\sigma \in C_{P_1}$ there exists $\sigma' \in C_{P_2}$ such that $\sigma \subseteq \sigma'$.*

Each node in the lattice corresponds to a simplet, with the root being the simplet that consists of a 0-simplex $\{r\}$, i.e., $O := (r, [r])$. A directed edge connects a node $A$ to a node $B$ iff $A \leq_C B$, and so we say that *A precedes B* in the lattice. The lattice can be generated by recursively applying the following expansion rules to a simplet $P$, initially set equal to the root:

**widen:** insert a new 1-simplex $[u, v]$ in $C_P$, such that $v \in V_P$ and $u \notin V_P$. This rule creates the simplet $P' := (V_P \cup \{u\}, C_P \cup \{[u, v]\})$, and hence increases the size of the vertex set $V_P$;

**inflate:** insert a new $q$-simplex $\sigma^{(q)}$ in $C_P$, such that its joist $J_{\sigma^{(q)}}$ is already contained in $C_P$. This rule creates the simplet $P' := (V_P, C_P \cup \{\sigma^{(q)}\})$, and hence may increase the dimension of $P$.[1]

Figure 2 illustrates a portion of the lattice. The applications of **widen** are denoted by green solid arrows, while the applications of **inflate** are denoted by blue dotted arrows. At every use of **widen**, we add a vertex to the simplet; while at every use of **inflate**, we strengthen the relations between the simplet vertices. In the latter case, the dimension of the simplet does not necessarily increase. For example, the first use of **inflate** generates the open triangle, which has the same dimension as the wedge (it contains only 1-simplices), while the second use of **inflate** generates the closed triangle, which, instead, has dimension 2.

The following theorem proves that the recursive application of the two expansion rules generates all the possible simplets. The proof can be found in Appendix A.2.

THEOREM 4.1 (COMPLETENESS). *Starting from a 0-simplex, the recursive application of **widen** and **inflate** can generate any simplet.*

Several nodes may precede the same node $A$, due to the existence of different pairs of simplices $(\sigma, \sigma')$ such that $\sigma \subset \sigma'$. In this case, node $A$ can be reached from the root via different paths. This situation corresponds to the existence of different sequences of applications of the two expansion rules that result in the same simplet. To avoid generating (and examining) the same simplet multiple times, we need to detect if a simplet has already been generated, i.e., we need to determine if two simplets are equal. In graph pattern mining, this is accomplished by either isomorphism testing or canonization. The canonization of a graph $G$ identifies a permutation of its edges that puts $G$ into its *canonical form* $\mathbb{C}(G)$, which satisfies the property that $\mathbb{C}(G) = \mathbb{C}(G')$ iff $G$ is isomorphic to $G'$. Similarly, we want to find the canonical form of the simplets.

We recall that simplets are complexes, and as such, they are hypergraphs satisfying the downward closure property. While the general hypergraph isomorphism problem has complexity $2^{O(n)}$, Arvind et al. [5] showed that there is no canonization procedure faster than $n!$, and proposed a canonization algorithm with complexity $(n+m)^{O(n)}$. Despite finding the canonical form of a simplet might be harder than testing isomorphism, thanks to the canonical forms of the simplets we can check all the equalities in linear time, and avoid checking isomorphism between each pair of simplets.

Since finding the canonical form of a graph is faster than finding the canonical form of a simplet [5], we follow a **delayed canonization** approach that tests several conditions of increasing complexity, and finds the canonical form of the simplet only if all the conditions hold. The first 4 conditions can be checked in linear time, while we can leverage the existing literature on graph isomorphism for the last condition [29]. Let $sc_P = [sc_{P,0}, \ldots, sc_{P,dim(P)}]$ be the *simplex dimension sequence* of $P$, with $sc_{P,i} = k$ iff $C_P$ contains exactly $k$ $i$-simplices; and $vc_P = [vc_{P,0}, \ldots, vc_{P,|V_P|}]$ be the *vertex degree sequence* of $P$, with $vc_{P,i} = k$ iff there exist exactly $k$ simplices in $C_P$ that contain vertex $v_i$. Then, given two simplets $P_1$ and $P_2$, if one of the following conditions does not hold, then $P_1 \not\equiv P_2$:

$$dim(P_1) = dim(P_2) \qquad |C_{P_1}| = |C_{P_2}| \qquad (1)$$
$$sc_{P_1} = sc_{P_2} \qquad sorted(vc_{P_1}) = sorted(vc_{P_2}) \qquad (2)$$
$$1\text{-}skeleton(P_1) \simeq 1\text{-}skeleton(P_2) \qquad (3)$$

Equations 1 state that the simplets must have the same dimension and number of simplices; Equations 2 that the dimensions of the simplices in the simplets must be the same, and that the vertices of the simplets must be part of the same number of simplices;[2] and, Equation 3 that the simplets must have the same 1-skeleton.

If all the conditions hold, we need to compare the canonical forms of the two simplets. Our solution to find those canonical forms is to reduce the problem to a graph canonization problem. We construct a polynomial-time reduction from the simplet to a bipartite graph, such that the left-hand side vertices are the vertices in the simplet, the right-hand side vertices are the simplices in the simplet, and an edge indicates the membership of a simplet vertex to a simplex. While for hypergraphs this reduction can exponentially increase the input size (the number of possible edges in a hypergraph with $n$ vertices is $2^n$), a reduction from a complex is far more feasible, thanks to the downward closure property of the complexes. In fact, for a complex with $n$ vertices, the number of vertices of the bipartite graph is upper-bounded by $n(n+1)/2$, and the number of edges is upper-bounded by $n(n-1)$. Therefore, we find the canonical form of a complex by first mapping the complex to a bipartite graph, and then finding the canonical form of the bipartite graph via a state-of-the-art library for graph isomorphism [29].

---

[1] Note that a pair of vertices $(u, v)$ in $P$ that are not faces of a common $q$-simplex in $C_P$ is the open joist $J_\sigma = \{[u], [v]\}$ of the 1-simplex $\sigma = [u, v]$.

[2] Note that the two vertex degree sequences are sorted in increasing order.

# 5 ALGORITHM

FreSCo (Frequent Simplets in a Complex, Algorithm 1) extracts the frequent simplets in a simplicial complex, given a maximum size $s^*$ and a minimum dimension $d^*$, simplet-growth approach that starts from $P := (\{v\}, \{[v]\})$, and recursively expands each frequent simplet until it becomes infrequent (Procedure EXPAND). For each candidate frequent simplet, FreSCo searches for subcomplexes in the complex that are isomorphic to the simplet, and stores the corresponding mapping from complex vertices to simplet vertices (Procedure EXAMINE-DP). Once it has found enough matches, it computes the support of the simplet, and if it is larger than the minimum frequency threshold $\tau$, the simplet is further expanded. If needed, FreSCo finds the canonical form of the expanded simplets to determine if they are redundant and should be discarded. The algorithm exploits several heuristics to guide the search of the occurrences and hence reduce the time required to assess the simplet's frequency. We next describe the algorithm that answers the decision problem introduced in Section 3. A variant of FreSCo that finds the exact frequency of the simplets, at the cost of higher running time is detailed in Appendix A.1.

**Algorithm 1** FreSCo

**Require:** Complex $\mathcal{K}$, Frequency Threshold $\tau$, Max Size $s^*$, Min Dim $d^*$
**Ensure:** Set of simplets $P$ with $SUP(P, \mathcal{K}) \geq \tau$
1: $O \leftarrow (\{v\}, \{[v]\}); I_O(v) \leftarrow V_\mathcal{K}$
2: $\mathcal{F} \leftarrow \text{EXPAND}(\mathcal{K}, O, \tau, s^*, d^*)$
3: **return** $\mathcal{F}$
4: **function** EXPAND($\mathcal{K}, P, \tau, s^*, d^*$)
5:    $E \leftarrow \emptyset; F \leftarrow \emptyset$
6:    **if** $|V_P| < s^*$ **then**
7:       **for** $v \in V_P$ **do**
8:          $u \leftarrow$ VERTEX()
9:          $P' \leftarrow (V_P \cup \{u\}, C_P \cup \{[u, v]\})$ ▷ widen
10:         $I_{P'}(w) \leftarrow V_\mathcal{K}$ for $w \in V_{P'}$
11:         **if** $P'$ not yet generated **then** $E \leftarrow E \cup \{(P', I_{P'})\}$
12:    **for** $J_{\sigma^{(q)}}$ such that $\sigma^{(q)} \notin C_P$ **do**
13:       $P' \leftarrow (V_P, C_P \cup \{\sigma^{(q)}\})$ ▷ inflate
14:       $I_{P'}(w) \leftarrow V_\mathcal{K}$ for $w \in V_{P'}$
15:       **if** $P'$ not yet generated **then** $E \leftarrow E \cup \{(P', I_{P'})\}$
16:    **for** $P, UI \in E$ **do**
17:       $I_P \leftarrow$ EXAMINE($\mathcal{K}, P, UI, \tau$); $sup \leftarrow \min_{v \in I_P} |I_P(v)|$
18:       **if** $sup \geq \tau$ **then**
19:          **if** $dim(P) \geq d^*$ **then** $F \leftarrow F \cup \{P\}$
20:          $F \leftarrow F \cup$ EXPAND($\mathcal{K}, P, \tau, s^*, d^*$)
21:    **return** $F$

**Simplet expansion.** The computation starts from a simplet $O$ consisting of a single vertex $v$, whose support set $I_O(v)$ is the vertex set of the complex. The set of frequent simplets $\mathcal{F}$ is initialized as the empty set. The main body of Algorithm 1 lies in Procedure EXPAND, which is called recursively to expand simplets $P$ proven to be frequent. If the number of vertices in $P$ is below the maximum size $s^*$, the simplet is first expanded by adding a new vertex $u$ and connecting it to a vertex in $V_P$ (**widen**). The algorithm applies **inflate** in the for loop at lines 12-15. For each joist $J_\sigma^{(q)}$ of a simplex $\sigma^{(q)}$ not already in the simplet $P$, a new simplet $P'$ is instantiated by adding $\sigma^{(q)}$ to $C_P$. For each expansion $P'$, the image sets of the vertices in $V_{P'}$ are initialized with $V_\mathcal{K}$. To identify the joists in the simplet we adopt a technique similar to the one in [41]: a dynamic inverted index that stores, for each face, all the simplices in the simplet that share that face. Since all the simplices in a joist must share a face by definition, this index allows us to speed up the search of the joists. In particular, for each $q$-simplex $\sigma$ in the simplet, we search for a subset $S$ of other simplices associated to the same face that share a vertex $w$ not in $\sigma$. If $|S| = q + 1$ and the simplex $\sigma \cup \{w\}$ does not exist in the simplet, $S \cup \{\sigma\}$ is a joist.

The same simplet $P'$ can be generated by applying different sequences of **widen** and **inflate**. Therefore, before adding $P'$ to $E$ (lines 11 and 15), we check if $P'$ has already been generated. We use a dictionary with key $(sc_P, vc_P)$ (i.e., simplicial dimension and simplicial degree sequence) to store the distinct simplets $P$ generated so far. Then, we perform the checks in Equations 1-3 with respect to each $P$ in the dictionary with the same key as $P'$. If all the checks are satisfied for some $P$, we find the canonical form of $P'$ and compare it to the canonical form of $P$. To speed up the computation of the canonical form, we employ a compact simplet representation such that $C_{P'}$ includes only maximal simplices.

Once the algorithm has created all the expansions $P'$ of $P$, the image sets are computed by Procedure EXAMINE-DP in Algorithm 2 (Appendix A.1). Then, if the minimum size of the image sets of $P'$ is above the threshold $\tau$, $P'$ is further expanded. The simplet is added to the output set $F$ only if its dimension $dim(P')$ is greater than the minimum threshold $d^*$. Retaining only the simplets of higher dimension leads to a more compact output set, and thus simplifies its examination by the end user. The full set of frequent simplets can still be obtained by setting $d^* = 1$.

**Frequency computation.** In the following, we say that two vertices are *neighbors* if they belong to a common simplex, and denote with $\Gamma(u)$ the set of neighbors of a vertex $u$. Algorithm 2, whose pseudocode is in Appendix A.1, searches for assignments from vertices in the complex to vertices in $P$, such that the membership relations to the simplices in $C_P$, are preserved.

The algorithm takes in input a parameter ($t^*$) for the maximum amount of time that the algorithm can spend trying to find a complete valid assignment for a candidate $u$, and the parent simplet of $P$, denoted with $SP$. The parent is the simplet that we expanded to generate $P$. The threshold $t^*$ is needed to avoid spending too much time on problematic candidates. The algorithm looks for *valid* assignments among the vertices in the sets $UI$, initialized with the vertex set of $\mathcal{K}$. We initialize the set of valid assignments $I_P(v)$ with the set of all the vertices of simplices having the same or more vertices than $P$, as all the simplices with $d$ vertices are sub-simplets of simplices with at least $d$ vertices. The sets of *non candidate* complex vertices $SP.NC$ are used to initialize the non-candidate sets $NC$ of $P$, as those vertices cannot be valid assignments for $P$ either.

The algorithm iterates over the vertices of $P$ examining only those for which it has not found enough assignments yet. When examining a vertex $v$, the set of candidates *Cands* is initialized as the set $V_\mathcal{K}$ minus $NC(v) \cup I_P(v)$, as these vertices cannot increase the size of $I_P(v)$, and hence do not help in determining if $P$ is frequent. To speed up the search of the valid assignments, *Cands* is sorted so that we examine first the candidates that were valid assignments for the vertex $v$ of $SP$. For each candidate $u$ that has at least the same number of neighbors of $v$, we initialize a dictionary $M$ with $M[v] = u$. The dictionary, which stores the partial vertex assignment, is expanded by Procedure FINDMATCH, which searches for valid assignments for all the other vertices in the simplet. If a complete assignment cannot be found, the vertex $u$ is added to $NC(v)$. Moreover, two conditions allow us to early-stop the examination of $v$: when the remaining candidates are not enough to assess that the simplet is frequent, and when $|I_P(v)| \geq \tau$.

Procedure findMatch recursively visits all the not-yet-explored vertices, according to an order obtained by a d.f.s. starting from $v$, so that the recursive call visits neighbor vertices most of the times. When visiting vertex $x$, the algorithm first computes a restricted candidate set $Cands$, and then tries to find a valid assignment for $x$ among the complex vertices in $Cands$. To create the restricted candidate set, we first compute the intersection among the neighbors of the complex vertices associated to simplet vertices that are neighbors of $x$. Thus, we can avoid examining complex vertices that would not preserve the simplex membership relation. A vertex assignment is *valid* if Procedure satisfiesConstraints returns true. If a candidate $n$ is a valid vertex assignment, the assignment is added to $M$, and Procedure findMatch is recursively called. For each simplex $\sigma$ in the simplet containing the vertex $x$, Procedure satisfiesConstraints checks whether the complex vertices already assigned to the vertices in $\sigma$ form, together with the candidate assignment $n$ of $x$, a simplex in the complex. If this is the case for each $\sigma$, the procedure returns true.

If a complete assignment starting from $u$ cannot be found in $t^*$ time, Procedure findMatch returns the empty set. This way, Algorithm 2 knows that it must store $u$ in the set *resume*, which includes vertices that are reexamined later, if $|I_P(v)| < \tau$.

At the end of Procedure findMatch, $M$ contains an assignment found starting with $M[v] = n$. If the assignment is complete, the image sets $I_P$ are updated with $M$, by exploiting the *orbit* information. Given a simplet $P$ and a vertex $v \in V_P$, the *orbit* $\Omega(v, P)$ of $v$ in $P$ is the subset of $V_P$ mapped to $v$ by any automorphism of $P$, i.e., $\Omega(v, P) = \{u \in V_P \mid \exists \phi \in Auto(P) . \phi(u) = v\}$. By definition, vertices in the same orbit have the same image set. Therefore, we propagate the valid assignments in $M$ to vertices in the same orbit.

## 6 EXPERIMENTAL EVALUATION

FreSCo code, implemented in Java 1.8, is available at https://github.com/lady-bluecopper/FreSCo together with the 6 simplicial complexes used in the experiments (Table 2 in Appendix A.3). Experiments are run on a 24-Core (1.90 GHz) Intel(R) Xeon(R) E5-2420 with 126GB of RAM, limiting the memory to 110GB, and using all the cores. We set $s^* = 5$ and $d^* = 1$ for all the experiments.

**Scalability.** We next evaluate the scalability of FreSCo (decision version). Figure 3 shows the running time of FreSCo for 5 increasing frequency thresholds. For each threshold, the algorithm finds roughly the same number of simplets in each dataset. The running time follows a concave curve: at higher frequencies there are few candidate frequent simplets, and therefore the algorithm terminates earlier; while at lower frequencies, even though the number of candidate frequent simplets is much larger, the presence of higher-order simplices in the complex simplifies the task of searching for occurrences. Moreover, by comparing the running times in DBLP with those in Enron, we can see that a 30× increase in dataset size leads to a time increase of only one order of magnitude.

**Comparison with graph pattern mining.** To assess the importance of mining higher-order structures when the input is a simplicial complex, Figure 4 presents an overview of the frequent simplets found at different frequency levels in Zebra (left) and DBLP (middle). These charts illustrate the number of frequent simplets for each dimension. If we modeled the datasets as simple graphs, we would find only the simplets denoted in dark green, corresponding to 1-dimensional structures which can be mapped to graph edges without loss. By looking for frequent simplets in the datasets modeled as simplicial complexes, we are able to detect a large number of higher-order structures. In particular, in orange we can see the number of frequent simplets consisting of simplices with dimension at most 2 (i.e., closed triangles), while the light blue bar indicates the number of simplets including simplices with dimension at most 3 (i.e., closed tetrahedra). Some of the simplets of dimension 3 persist over multiple frequency thresholds, which indicates that these higher-order structures are not trivially frequent. All the simplets with dimension larger than 1 can be neither modeled nor mined by adopting traditional graph-based approaches. In particular, these algorithms cannot discriminate between the $i$-skeleton with $i \leq q$ of a simplet of dimension $q$, meaning that the occurrences of a closed triangles are treated as occurrences of an open triangle.

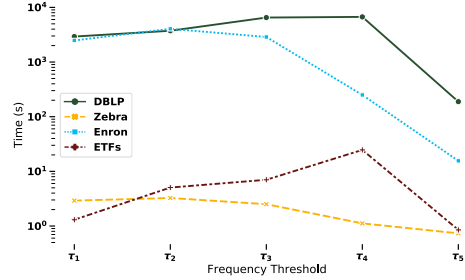

Figure 3: Scalability in various simplicial complexes.

Finally, we observe that in Zebra the simplets are concentrated in the small frequency range $[4200, 4800]$, which corresponds to $[41\%, 47\%]$ of the complex size; while in DBLP, they appear in the larger frequency range $[870k, 1.5M]$, which corresponds to $[45\%, 78\%]$ of the complex size. The fact that frequent simplets exists at larger frequencies indicates that the entities in DBLP are more interconnected than in Zebra, as there are several researchers that collaborate in large groups. The differences in the distributions of frequent simplets highlight that the our proposal captures different characteristics of the underlying simplicial complexes.

**Exact frequency values.** Finding all the occurrences of a simplet in a complex is an expensive task (see Section 5). As an example, finding the exact frequencies in ETFs takes 160× more than the decision version of the problem; while in Zebra it takes 152× more. Trading exactness for shorter runtime might be meaningful in many applications, however, when computing the exact frequency values is feasible, we can use them to find the top-$k$ simplets. This filtering allow the user to reduce the size of the output, and get only the most relevant structures. We order all the simplets found in ETFs and Zebra, and compare their top-5 simplets to determine if they are characterized by different structures. Zebra is characterized by $P_1 : (\{0, 1, 2, 3\}, \{[0, 1, 3], [0, 1, 2]\})$, i.e., two closed triangles sharing one edge. ETFs is characterized by $P_2 : (\{0, 1, 2, 3, 4\}, \{[0, 3] - [0, 1, 2] - [1, 4]\})$, i.e., a closed triangle with two tails. In the literature, frequent patterns have been successfully used to learn vector representations of graphs for classification purposes [4]. Similarly, frequent simplets can be used to learn finer-grained vectors for complexes.

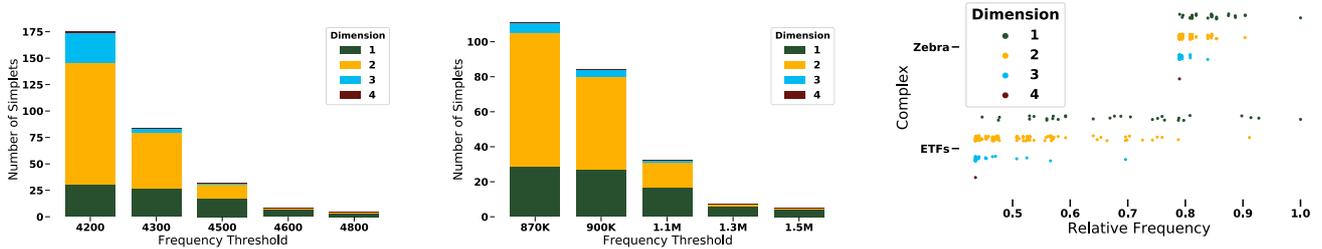

Figure 4: Number of frequent simplets of each dimension (indicated by color) found in Zebra (left) and DBLP (middle), when varying the frequency threshold. Relative frequencies of the simplets w.r.t. the most frequent one in Zebra and ETFs (right).

Figure 4 (right) compares the distributions of frequent simplets in the two complexes. For each simplet of each dimension, we report the relative frequency with respect to the largest frequency in the complex. From this chart, we can better understand how close are the frequent simplets in Zebra, with all the relative frequencies being higher than 0.78. In contrast, in ETFs most of the simplets have less than 0.70 of the frequency of the top simplet, and hence the latter can be a good discriminator for this complex.

**Face-to-face interactions.** Studying face-to-face interactions in public places can help understanding how, e.g., transmissible diseases propagate. Each simplet encodes a different type of social relation, and frequent simplets indicate relations that happen frequently over time. We can focus on a specific simplet, and study how its frequency changes over time. Persistent simplets of higher dimension indicate patterns of communication with higher risk, as they represent an interaction among more parties [26] that took place frequently over time. These patterns cannot be effectively modeled via graph patterns: an open triangle in a simplicial complex indicates people that have pairwise contacts, while a closed triangle indicates three people that interact simultaneously. The two situations are not distinguishable in simple graphs, although the latter represents a situation of higher risk.

Figure 5 reports the number of frequent ($\tau = 40$) simplets found in the High (left) and the Primary (middle) temporal complexes (datasets description in Appendix A.3), with different colors indicating simplets of different dimension. Primary-school students interact more consistently over time, with a higher spike characterized by stronger contacts during the lunch break. In contrast, high-school students have fewer interactions in class (we observe only frequent simplets of dimension 1), and engage with other students mainly during lunch and before class.

Figure 6 shows the normalized entropy of the genders and classes in the occurrences of all the frequent simplets of dimension > 1, found in High (top) and Primary (bottom). Gender (Class) entropy indicates the tendency of people to gather with people of different gender (class). Therefore, smaller values of gender entropy indicate more homophilic interactions; while larger values denote heterophilic interactions. Primary-school students frequently interact with both female and male students, but during lunch, they tend to group together with children of the same gender. The heterophilic interactions take place more often during class, as classes consist of children of different genders. The dimensionality of these interactions is lower, and may be due to children sitting in adjacent desks. Similarly, high-school students interact more frequently with students of the same gender before class and during lunch, and hence the gender entropy is lower around 9am and 12am. By looking at the number of frequent simplets in the two charts, we note that primary-school students have stronger interactions during lunch. The values of class entropy tell us that students frequently mix with students of other classes, whenever they have the chance (e.g., during lunch and before/after class).

Finally, Figure 5 (right) reports the persistence of the frequent simplets of dimension 2 and 3, across all the snapshots of High and Primary. The x-axis is sorted by dimensionality of the simplet, i.e., the 3-dim simplets are on the right. In High, the simplets are less persistent, which indicates that high-school students communicate in large groups less frequently. The difference with Primary is more evident in the simplets of higher dimension. Few higher-order interactions persist among high-school students, whereas primary-school children meet with other children more consistently.

**Simplicial closure prediction.** Link prediction has the goal of predicting whether two nodes not connected in a network will be linked in the future, and is important in many applications like recommender systems [3]. Benson et al. [9] are the first to study the problem of link prediction in higher-order networks. In their seminal work, they extend the theory of triadic closure in social networks by defining the concept of simplicial closure, i.e., the tendency of groups of nodes with strong past interactions to co-appear in higher-order structures in the future. They focus on a specific case of simplicial closure, i.e., given a 3-node joist, predict whether the corresponding simplex will appear in the future. This problem is ignored in traditional link prediction, as the triangle would be already considered part of the graph. They evaluate the prediction performance of several models as the ratio between the area under the precision-recall curve (AUC-PR) and a random score, calculated as the proportion of open triangles in the training set that close in the test set. The models considered are harmonic (hm), geometric (gm), and arithmetic (am) means of the edge weights in the open triangle, Adamic-Adar coefficient (aa), preferential attachment (pa), personalized PageRank similarity (ppr), Katz similarity (ka), and logistic regression on 6 types of features (lr). The first 7 models provide a score for each triangle, while the latter is a supervised algorithm. Experimental results prove that the task is challenging, and that tie strength is important in predicting simplicial closure.

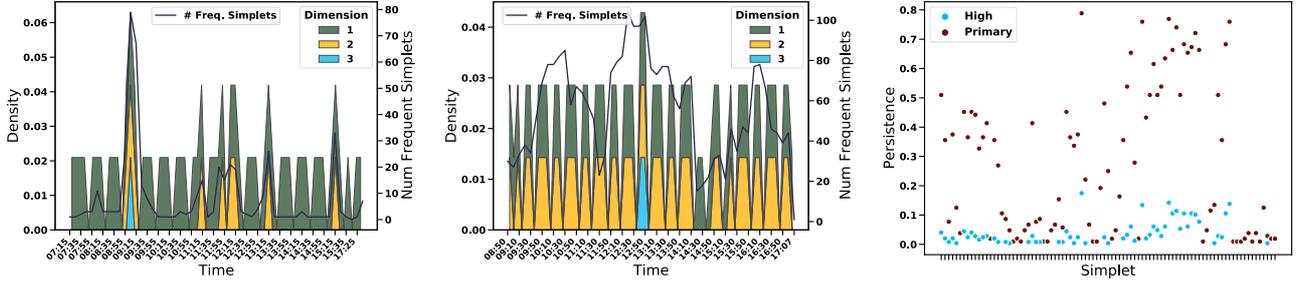

Figure 5: Number of frequent simplets found in the High (left) and the Primary (middle) temporal complexes, together with their persistence over the snapshots (right).

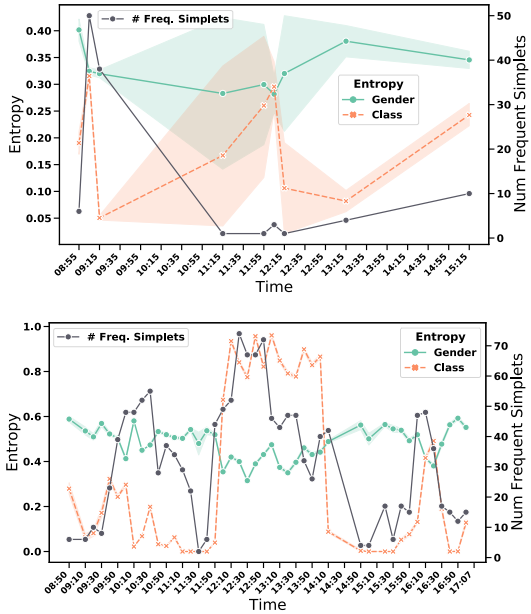

Figure 6: Gender and class entropy in the occurrences of the frequent simplets found in High (top) and Primary (bottom).

In our experiment, we consider the two temporal networks High and Primary, and focus on the task of predicting which open triangle at time $t$ will close at time $t + 1$. We generate our ground truth considering all the open triangles at time $t$, and assigning them 0 if they are open at time $t + 1$, and 1 otherwise. For each pair of vertices $e = (u, v)$ and each timestamp $t$, we create a binary vector $\mathbf{b}(e, t)$ such that $b(e, t)[i] = 1$ if $v$ appears in some occurrence of the $i$-th frequent simplet, and 0 otherwise. Then, we generate a feature vector for each triangle $T$ and each timestamp $t$, by taking the harmonic mean of the vectors $\mathbf{b}(e, t)$ of its edges.

We split the data into train and test set ($\frac{2}{3}$, $\frac{1}{3}$) randomly, and train a Gradient Boosting Classifier (GBC). Since the data is highly imbalanced towards class 0, we use class weights. We evaluate the model in terms of the normalized AUC-PR, averaged over 10 runs, and compare with the models considered by Benson [9]. Due to space limitations, we refer to [9] and the corresponding Supplementary Material[3] for their detailed description. Table 1 reports the normalized AUC-PR scores for all the models in High (H) and Primary (P).

[3]https://www.pnas.org/content/suppl/2018/11/08/1800683115.DCSupplemental

We observe that all the models achieve better performance than random guessing, with some models achieving one order of magnitude improvement. The fact that the mean-based models perform better than the other baselines indicate that the presence of many higher-order structures around an edge is a good predictor. The logistic regressor, which uses properties of the nodes in the projected graph and information about the simplices containing the nodes, overtakes all the baselines, as it exploits the training data to learn a better model. Our approach, GBC-FreSCo, achieves the best performance in Primary, and ranks third in High, proving that encoding the neighborhood of the edges in terms of the frequent simplets is a good predictor of triangle closure. The frequency threshold $\tau$ used to mine the simplets in the complex, considerably affects the performance of GBC-FreSCo, and hence must be properly tuned for the dataset at hand. We plan to further investigate the ability of frequent simplets to predict simplicial closure, by considering complexes of different nature and different frequency thresholds.

Table 1: Normalized AUC-PR scores in the triangle closure task in Primary (P) and High (H), for different models.

| $\mathcal{K}$ | hm | gm | am | aa | pa | ka | ppr | lr | GBC |
|---|---|---|---|---|---|---|---|---|---|
| P | 31.26 | 31.73 | 26.03 | 5.03 | 2.03 | 1.78 | 2.29 | 33.04 | **37.46** |
| H | 47.92 | 48.44 | 38.32 | 9.90 | 4.28 | 7.79 | 3.83 | **60.85** | 48.22 |

## 7 CONCLUSIONS

We formulated the problem of frequent simplet mining in simplicial complexes, which extends the traditional graph pattern mining to higher dimensions. We showed how to measure the frequency of a simplet in a complex, and how to generate all the candidate simplets up to a max size. We presented our algorithm, FreSCo, that exploits the relations between the simplices in the complex and between the simplets to speed up the search and prune the search space. In addition, FreSCo adopts a delayed canonization procedure, to reduce the amount of isomorphism checks between simplets. This way, it can scale with the size of the simplets. FreSCo can both answer the question "is a simplet $P$ frequent?", and find the exact frequency of the simplets. Our experimental evaluation showed that FreSCo can efficiently mine simplets in various complexes, and that simplets allow finer-grained characterizations of the complexes. Finally, we studied how relations change in two temporal complexes, and discussed the persistence and entropy of those relations.

# REFERENCES

[1] Ehab Abdelhamid, Ibrahim Abdelaziz, Panos Kalnis, Zuhair Khayyat, and Fuad Jamour. 2016. ScaleMine: Scalable parallel frequent subgraph mining in a single large graph. In *SC*. 716–727.

[2] Rakesh Agrawal and Ramakrishnan Srikant. 1994. Fast Algorithms for Mining Association Rules in Large Databases. In *VLDB*. 487–499.

[3] Luca Maria Aiello, Alain Barrat, Rossano Schifanella, Ciro Cattuto, Benjamin Markines, and Filippo Menczer. 2012. Friendship prediction and homophily in social media. *TWEB* 6, 2 (2012), 1–33.

[4] Md Alam, Chowdhury Farhan Ahmed, Md Samiullah, Carson K Leung, et al. 2021. Discriminating frequent pattern based supervised graph embedding for classification. In *PAKDD*. 16–28.

[5] Vikraman Arvind, Bireswar Das, Johannes Köbler, and Seinosuke Toda. 2010. Colored hypergraph isomorphism is fixed parameter tractable. In *FSTTCS*. Schloss Dagstuhl-Leibniz-Zentrum fuer Informatik.

[6] Cigdem Aslay, Muhammad Anis Uddin Nasir, Gianmarco De Francisci Morales, and Aristides Gionis. 2018. Mining Frequent Patterns in Evolving Graphs. In *CIKM*. 923–932.

[7] Ron Atkin. 1974. *Mathematical structure in human affairs*. Heinemann Educational Publishers.

[8] Austin Benson, David F. Gleich, and Jure Leskovec. 2016. Higher-order organization of complex networks. *Science* 353, 6295 (2016), 163–166.

[9] Austin R Benson, Rediet Abebe, Michael T Schaub, Ali Jadbabaie, and Jon Kleinberg. 2018. Simplicial closure and higher-order link prediction. *PNAS* 115, 48 (2018), E11221–E11230.

[10] Austin R. Benson, David F. Gleich, and Desmond J. Higham. 2021. Higher-order Network Analysis Takes Off, Fueled by Classical Ideas and New Data. *SIAM News* (2021).

[11] Michele Berlingerio, Fabio Pinelli, and Francesco Calabrese. 2013. ABACUS: frequent pAttern mining-BAsed Community discovery in mUltidimensional networkS. *Data Min. Knowl. Discov.* 27, 3 (2013), 294–320.

[12] Jacob Charles Wright Billings, Mirko Hu, Giulia Lerda, Alexey N Medvedev, Francesco Mottes, Adrian Onicas, Andrea Santoro, and Giovanni Petri. 2019. Simplex2Vec embeddings for community detection in simplicial complexes. *arXiv preprint arXiv:1906.09068* (2019).

[13] Björn Bringmann, Michele Berlingerio, Francesco Bonchi, and Aristides Gionis. 2010. Learning and Predicting the Evolution of Social Networks. *IEEE Intell. Syst.* 25, 4 (2010), 26–35.

[14] Björn Bringmann and Siegfried Nijssen. 2008. What is frequent in a single graph?. In *PAKDD*. 858–863.

[15] Yuri Dabaghian, Facundo Mémoli, Loren Frank, and Gunnar Carlsson. 2012. A topological paradigm for hippocampal spatial map formation using persistent homology. *PLoS Computational Biology* 8, 8 (2012).

[16] Prasanna Desikan and Jaideep Srivastava. 2006. Mining Temporally Changing Web Usage Graphs. In *Advances in Web Mining and Web Usage Analysis*, Bamshad Mobasher, Olfa Nasraoui, Bing Liu, and Brij Masand (Eds.). Springer Berlin Heidelberg, 1–17.

[17] W Dörfler and DA Waller. 1980. A category-theoretical approach to hypergraphs. *Archiv der Mathematik* 34, 1 (1980), 185–192.

[18] M. Elseidy, E. Abdelhamid, S. Skiadopoulos, and P. Kalnis. 2014. Grami: Frequent subgraph and pattern mining in a single large graph. *PVLDB* 7, 7 (2014), 517–528.

[19] Dhivya Eswaran, Srijan Kumar, and Christos Faloutsos. 2020. Higher-Order Label Homogeneity and Spreading in Graphs. In *The Web Conference 2020*. 2493–2499.

[20] M. Fiedler and C. Borgelt. 2007. Subgraph support in a single large graph. In *ICDMW*. 399–404.

[21] Valerio Gemmetto, Alain Barrat, and Ciro Cattuto. 2014. Mitigation of infectious disease at school: targeted class closure vs school closure. *BMC infectious diseases* 14, 1 (2014), 1–10.

[22] Vitali Herrera-Semenets, Niusvel Acosta-Mendoza, and Andrés Gago-Alonso. 2015. A Framework for intrusion detection based on frequent subgraph mining. In *Proceedings of the 2nd SDM Workshop on Mining Networks and Graphs: A Big Data Analytic Challenge*.

[23] Mehdi Heydari, Raed Ali Helal, and Khairil Imran Ghauth. 2009. A graph-based web usage mining method considering client side data. In *2009 International Conference on Electrical Engineering and Informatics*, Vol. 1. IEEE, 147–153.

[24] Danijela Horak, Slobodan Maletić, and Milan Rajković. 2009. Persistent homology of complex networks. *Journal of Statistical Mechanics: Theory and Experiment* 2009, 03 (2009), P03034.

[25] Tamás Horváth, Björn Bringmann, and Luc De Raedt. 2006. Frequent hypergraph mining. In *International Conference on Inductive Logic Programming*. 244–259.

[26] Iacopo Iacopini, Giovanni Petri, Alain Barrat, and Vito Latora. 2019. Simplicial models of social contagion. *Nature Communications* 10, 1 (2019), 1–9.

[27] Kasra Jamshidi, Rakesh Mahadasa, and Keval Vora. 2020. Peregrine: a pattern-aware graph mining system. In *EuroSys*. 1–16.

[28] Ruoming Jin and G. Agrawal. 2005. An algorithm for in-core frequent itemset mining on streaming data. In *ICDM*. 8–pp.

[29] T. Junttila and P. Kaski. 2007. Engineering an efficient canonical labeling tool for large and sparse graphs. In *ALENEX*. 135–149.

[30] Xiao Liu, Nanyin Zhang, Catie Chang, and Jeff H Duyn. 2018. Co-activation patterns in resting-state fMRI signals. *Neuroimage* 180 (2018), 485–494.

[31] Louis-David Lord, Paul Expert, Henrique M. Fernandes, Giovanni Petri, Tim J. Van Hartevelt, Francesco Vaccarino, Gustavo Deco, Federico Turkheimer, and Morten L. Kringelbach. 2016. Insights into Brain Architectures from the Homological Scaffolds of Functional Connectivity Networks. *Frontiers in Systems Neuroscience* 10 (2016), 85.

[32] Eugene M Luks. 1999. Hypergraph isomorphism and structural equivalence of boolean functions. In *ACM Symposium on Theory of Computing*. 652–658.

[33] Rossana Mastrandrea, Julie Fournet, and Alain Barrat. 2015. Contact patterns in a high school: a comparison between data collected using wearable sensors, contact diaries and friendship surveys. *PloS one* 10, 9 (2015), e0136497.

[34] Changping Meng, S Chandra Mouli, Bruno Ribeiro, and Jennifer Neville. 2018. Subgraph pattern neural networks for high-order graph evolution prediction. In *AAAI Conference on Artificial Intelligence*, Vol. 32.

[35] S. Moens, E. Aksehirli, and B. Goethals. 2013. Frequent Itemset Mining for Big Data. In *BigData*. 111–118.

[36] Muhammad Anis Uddin Nasir, Cigdem Aslay, Gianmarco De Francisci Morales, and Matteo Riondato. 2021. TipTap: Approximate mining of frequent k-Subgraph patterns in evolving graphs. *TKDD* 15, 3 (2021), 1–35.

[37] C. C. Noble and D. J. Cook. 2003. Graph-based Anomaly Detection. In *SIGKDD*. 631–636.

[38] Takeshi Obayashi, Yuki Kagaya, Yuichi Aoki, Shu Tadaka, and Kengo Kinoshita. 2018. COXPRESdb v7: a gene coexpression database for 11 animal species supported by 23 coexpression platforms for technical evaluation and evolutionary inference. *Nucleic Acids Research* 47, D1 (2018), D55–D62.

[39] Alice Patania, Giovanni Petri, and Francesco Vaccarino. 2017. The shape of collaborations. *EPJ Data Science* 6, 1 (2017), 1–16.

[40] Giovanni Petri, Paul Expert, Federico Turkheimer, Robin Carhart-Harris, David Nutt, Peter J Hellyer, and Francesco Vaccarino. 2014. Homological scaffolds of brain functional networks. *Journal of The Royal Society Interface* 11, 101 (2014), 20140873.

[41] Giulia Preti, Gianmarco De Francisci Morales, and Francesco Bonchi. 2021. STruD: Truss Decomposition of Simplicial Complexes. In *Proceedings of The Web Conference 2021*. 3408–3418.

[42] Giulia Preti, Gianmarco De Francisci Morales, and Matteo Riondato. 2021. MaNIACS: Approximate mining of frequent subgraph patterns through sampling. In *KDD*. 1348–1358.

[43] Giulia Preti, Matteo Lissandrini, Davide Mottin, and Yannis Velegrakis. 2018. Beyond Frequencies: Graph Pattern Mining in Multi-weighted Graphs.. In *EDBT*, Vol. 18. 169–180.

[44] Saif Ur Rehman and Sohail Asghar. 2020. Online social network trend discovery using frequent subgraph mining. *Social Network Analysis and Mining* 10, 1 (2020), 1–13.

[45] Vsevolod Salnikov, Daniele Cassese, and Renaud Lambiotte. 2018. Simplicial complexes and complex systems. *European Journal of Physics* 40, 1 (2018), 014001.

[46] Daniel Hernández Serrano and Darío Sánchez Gómez. 2019. Centrality measures in simplicial complexes: applications of TDA to Network Science. *arXiv preprint arXiv:1908.02967* (2019).

[47] David I Spivak. 2009. Higher-dimensional models of networks. *arXiv preprint arXiv:0909.4314* (2009).

[48] Leo Torres, Ann S Blevins, Danielle S Bassett, and Tina Eliassi-Rad. 2020. The why, how, and when of representations for complex systems. *arXiv preprint arXiv:2006.02870* (2020).

[49] Di Wu, Jiadong Ren, and Long Sheng. 2017. Uncertain maximal frequent subgraph mining algorithm based on adjacency matrix and weight. *International Journal of Machine Learning and Cybernetics* 9, 9 (2017), 1445–1455.

[50] Kelin Xia and Guo-Wei Wei. 2014. Persistent homology analysis of protein structure, flexibility, and folding. *International journal for numerical methods in biomedical engineering* 30, 8 (2014), 814–844.

[51] Jean-Gabriel Young, Giovanni Petri, Francesco Vaccarino, and Alice Patania. 2017. Construction of and efficient sampling from the simplicial configuration model. *Physical Review E* 96, 3 (2017), 032312.

## A SUPPLEMENTARY MATERIAL

In the following, we provide the pseudocode of the procedures used by FreSCo, the proofs of the main theorems, and characteristics and description of the datasets used in the experimental analysis.

---

**Algorithm 2** EXAMINE (DP)

**Require:** Complex $\mathcal{K}$, Simplet $P$, Parent simplet $SP$
**Require:** Frequency Threshold $\tau$, Timeout $t^*$
**Ensure:** Image sets $I_P$ of the vertices in $V_P$
1: $NC \leftarrow SP.NC$
2: $H \leftarrow \{v \in V_\mathcal{K} \mid \exists \sigma \in \mathcal{K} \land v \in \sigma \land |V_\sigma| \geq |V_P|\}$
3: $I_P(v) \leftarrow H$ for $v \in V_P$
4: **for** $v \in V_P$ **do**
5:   **if** $|I_P(v)| < \tau$ **then**
6:     $Cands \leftarrow V_\mathcal{K} \setminus (I_P(v) \cup NC(v))$
7:     sort $Cands$ placing first the vertices in $I_{SP}(v)$
8:     **if** $|Cands \cup I_P(v)| < \tau$ **then return** $\emptyset$
9:     $c \leftarrow 0$; $resume \leftarrow \emptyset$
10:     **for** $u \in Cands$ **do**
11:       $c \leftarrow c + 1$
12:       **if** $|\Gamma(n)| < |\Gamma(v)|$ **then continue**
13:       $M \leftarrow \emptyset$; $M[v] \leftarrow u$;
14:       $M \leftarrow \text{findMatch}(\mathcal{K}, P, M, time(), t^*)$
15:       **if** $|M| = |V_P|$ **then**
16:         update and propagate $I_P$ with $M$
17:       **else if** $M = \emptyset$ **then**
18:         $resume \leftarrow resume \cup \{u\}$; $c \leftarrow c - 1$
19:       **else** $NC(v) \leftarrow NC(v) \cup \{u\}$
20:       **if** $|Cands \cup I_P(v)| - c < \tau$ **then return** $\emptyset$
21:       **if** $|I_P(v)| \geq \tau$ **then break**
22:     **if** $|I_P(v)| < \tau$ **then** do loop at 10-21 for cands in $resume$
23: $P.NC \leftarrow NC$
24: **return** $I_P$

25: **function** findMatch($\mathcal{K}, P, M, t, t^*$)
26:   **if** $time() - t > t^*$ **then return** $\emptyset$
27:   **if** $|M| = |V_P|$ **then return** $M$
28:   $x \leftarrow$ next vertex to match
29:   $Cands \leftarrow \left(\bigcap_{w \in M. \; w \in \Gamma(x)} \Gamma(M[w])\right)$
30:   **for** $n \in Cands$ **do**
31:     **if** satisfiesConstraints($\mathcal{K}, P, M, x, n$) **then**
32:       $M[x] \leftarrow n$; $M \leftarrow$ findMatch($\mathcal{K}, P, M, t, t^*$)
33:       **if** $|M| = |V_P| \land M = \emptyset$ **then return** $M$
34:     **else if** $time() - t > t^*$ **then return** $\emptyset$
35:   **return** $M$

36: **function** satisfiesConstraints($\mathcal{K}, P, M, x, n$)
37:   **for** $\sigma \in C_P$ such that $x \in \sigma$ **do**
38:     **if** $M \restriction \sigma \cup \{n\} \notin \mathcal{K}$ **then return false**
39:   **return true**

---

### A.1 Pseudocode

We report the pseudocode of Procedure EXAMINE invoked at line 17 of Algorithm 1 in its two versions: Algorithm 2 determines if a simplet is frequent without computing the exact frequency, and Algorithm 3 finds the exact frequency of all the frequent simplets. We also provide a brief description of Algorithm 3.

**Computing exact frequencies.** Algorithm 3 provides the procedure EXAMINE to find the exact frequencies. To save time, the algorithm looks for *valid* assignments for a vertex $v$ in the expansion $P$, only among vertices that were valid assignments for the corresponding vertex in the parent simplet $SP$, i.e., the set of candidates $UI(v)$ is initialized with $I_{SP}(v)$. However, if $v$ is a newly added vertex, $UI(v) \leftarrow V_\mathcal{K}$. This initialization leverages that only valid occurrences of $SP$ can be expanded into valid occurrences of its expansions $P$. The algorithm sorts the vertices $v$ in the simplet by increasing size of $UI(v)$, with the hope to detect earlier if the simplet is not frequent, and hence stop the examination. Then, it iteratively visits each vertex $v$ to compute the image set $I_P(v)$. Given that an assignment $u$ may have already been added to $I_P(v)$ when visiting another simplet vertex, we initialize the set of candidates as the set difference between $UI(v)$ and $I_P(v)$. Similarly to Algorithm 2, once assigned a candidate $u$ to $v$, Procedure findMatches finds an assignment for all the other vertices in the simplet. Procedure findMatches examines each candidate match $n$ for each other vertex $x$ of $P$, and differs from Procedure findMatch in Algorithm 2, in two parts. First, it creates the restricted set $Cands$ via intersection with the upper bound set $UI(x)$ (line 18). Secondly, it has no time constraints, and hence searches for a valid assignment starting from $u$ until it either finds it or it has explored all the possibilities.

---

**Algorithm 3** EXAMINE (EF)

**Require:** Complex $\mathcal{K}$, Simplet $P$, Sets of Candidates $UI$
**Require:** Frequency Threshold $\tau$
**Ensure:** Image sets $I_P$ of the vertices in $V_P$
1: $H \leftarrow \{v \in V_\mathcal{K} \mid \exists \sigma \in \mathcal{K} \land v \in \sigma \land |V_\sigma| \geq |V_P|\}$
2: $I_P(v) \leftarrow H$ for $v \in V_P$
3: sort $v \in V_P$ by size of $UI(v)$
4: **for** $v \in V_P$ **do**
5:   $Cands \leftarrow UI(v) \setminus I_P(v)$; $c \leftarrow 0$
6:   **for** $u \in Cands$ **do**
7:     $c \leftarrow c + 1$
8:     **if** $|\Gamma(u)| < |\Gamma(v)|$ **then continue**
9:     $M \leftarrow \emptyset$; $M[v] \leftarrow u$
10:     $M \leftarrow \text{findMatches}(\mathcal{K}, P, UI, M)$
11:     **if** $|M| = |V_P|$ **then**
12:       update and propagate $I_P$ with $M$
13:     **if** $|Cands \cup I_P(v)| - c < \tau$ **then return** $\emptyset$
14: **return** $I_P$

15: **function** findMatches($\mathcal{K}, P, UI, M$)
16:   **if** $|M| = |V_P|$ **then return** $M$
17:   $x \leftarrow$ next vertex to match
18:   $Cands \leftarrow \left(\bigcap_{w \in M. \; w \in \Gamma(x)} \Gamma(M[w])\right) \cap UI(x)$
19:   **for** $n \in Cands$ **do**
20:     **if** satisfiesConstraints($\mathcal{K}, P, M, x, n$) **then**
21:       $M[x] \leftarrow n$; $M \leftarrow$ findMatches($\mathcal{K}, P, UI, M$)
22:       **if** $|M| = |V_P|$ **then return** $M$
23:   **return** $M$

---

### A.2 Proofs

**Proof of Theorem 4.1.**

PROOF. Let $P$ be a simplet and $\sigma^{(q)}$ be a $q$-simplex in $C_P$. By definition, $\sigma^{(q)}$ consists in a set of $q+1$ interconnected vertices. W.l.o.g., pick a vertex $v$ in $\sigma^{(q)}$ and start the procedure from $[v]$. By applying **widen** $q$ times, we add the remaining vertices in $\sigma^{(q)}$, connecting them to $[v]$ via 1-simplices. This generates a star structure with $[v]$ at the center. Then, by applying **inflate** $q(q-1)/2$ times, we add the remaining 1-simplices between the vertices, and by applying it other $\sum_{i=3}^{q+1} \binom{q+1}{i}$ times, we obtain $\sigma^{(q)}$. In particular, for each $i \geq 3$, the $\binom{q+1}{i}$ applications add all the $(i-1)$-simplices. By definition of simplet, all the simplices in $C_P$ are connected to at least another simplex in $C_P$. Therefore, there exists at least one $\sigma_\diamond^{(r)} \in C_P$ that shares a face with $\sigma^{(q)}$. By applying **widen**, we add to the simplet a vertex in $\sigma_\diamond^{(r)}$ but not in $\sigma^{(q)}$. Then, by applying **inflate** using the same procedure as before, we obtain the simplet $\widetilde{P}$ with $C_{\widetilde{P}} = \{\sigma^{(q)}, \sigma_\diamond^{(r)}\}$. By repeating the last two steps for each other simplex in $P$, we obtain $\widetilde{P} = P$. □

**Anti-monotonicity of** *SUP*. Proof of Property 1.

PROOF. We prove the anti-monotone property by contradiction. Let consider a simplet $P$ and its sub-simplet $SP$, and let $\psi$ be an injective function from $V_{SP}$ to $V_P$ that preserves the relations between the vertices (we can define such $\psi$ because $C_{SP} \subset C_P$). Let assume that $P$ has frequency $SUP(P, \mathcal{K})$ greater than $SP$. By definition of $SUP$, this means that for all $v \in V_{SP}$ such that $|I_{SP}(v)| = SUP(SP, \mathcal{K})$, it holds that $|I_P(\psi(v))| \geq |I_{SP}(v)|$. W.l.o.g. let us pick one simplet vertex $w = \psi(v)$ and a complex vertex $n \in I_P(w) \setminus I_{SP}(v)$. By definition of image set, if $n$ belongs to $I_P(w)$, then there exists a bijection $\phi$ from a subset of complex vertices to $V_P$ such that **(i)** $\phi(n) = w$ and **(ii)** it preserves the relations between the vertices. However, this means that $\vartheta = \psi^{-1} \circ \phi_{\restriction Cdm(\psi)}$, where $Cdm(\psi)$ is the codomain of $\psi$, is a valid bijection from a subset of complex vertices to $V_{SP}$ such that **(i)** $\vartheta(n) = v$ and **(ii)** it preserves the relations between the vertices. We reach a contradiction, because we assumed $n \in I_P(w) \setminus I_{SP}(v)$, but by definition of image set, we must have that $n \in I_{SP}(v)$. □

**Completeness and correctness of Algorithm 1.**

THEOREM A.1 (COMPLETENESS AND CORRECTNESS). *Algorithm 1 returns all and only the simplets $P$ with $SUP(P, \mathcal{K}) \geq \tau$.*

PROOF. Theorem 4.1 proves that the expansion strategy used by Procedure EXPAND generates the complete lattice. Thanks to the anti-monotonicity property of the *SUP* measure, a simplet can be frequent only if its sub-simplets are frequent. Therefore, by expanding only the frequent simplets we do not discard any candidate frequent simplet. This ensures the completeness of the algorithm.

We first show the correctness of Algorithm 2. This algorithm searches for an assignment for a vertex $v$ among all the complex vertices, and hence, if a valid assignment exists, it can find it. Early stops happen in two cases: if it has found at least $\tau$ assignments for all the vertices in $P$, or if the remaining candidates are not enough to prove that $P$ is frequent. In the first case, $SUP(P, \mathcal{K}) \geq \tau$ and hence $P$ is correctly added to the output set. In the second case, the simplet cannot be frequent, because $SUP(P, \mathcal{K})$ cannot be greater than $\tau$ even if all the remaining candidates are valid assignments, and hence we do not need to examine them. Therefore, the algorithm ensures a correct answer to the question $SUP(P, \mathcal{K}) \geq \tau$.

We now show the correctness of Algorithm 3. This algorithm initializes the sets of candidates for the simplet vertices of $P$ with the image sets of the parent simplet of $P$. This optimization leads to correct results, because only valid assignments for sub-simplets of $P$ can be expanded to valid assignments for $P$. The algorithm iteratively assigns each candidate $u$ to each simplet vertex $v$, and then tries to find an assignment for all the other simplet vertices via Procedure FINDMATCHES. Since FINDMATCHES examines each candidate for all the other simplet vertices, if a valid assignment exists, we surely find it. The validity of the assignment is checked in lines 18 and 38, where we ensure that assignments for vertices that are connected in the simplet are connected in the complex. Therefore, Algorithm 3 leads to the correct computation of $SUP(P, \mathcal{K})$, and hence ensures a correct answer to the question $SUP(P, \mathcal{K}) \geq \tau$. □

**Complexity of FRESCO** Reference [32] proves that the isomorphism problem for sub-hypergraphs is NP-hard. As complexes are special kinds of hypergraphs, the simplet isomorphism problem is also NP-hard. Let $S_k = \sum_{j=2}^{k} \binom{k}{j}$ denote the number of $(j-1)$-simplices with up to $k$ vertices. Then, the number of simplets with $k$ vertices is upper bounded by $2^{S_k}$ (there is no known formula for the exact number of distinct simplets). To determine the frequency of a simplet with $k$ vertices, we need all its isomorphisms in the complex, which takes up to $O(|V|^k)$. Therefore, examining all the simplets of size $k$ requires up to $O(2^{S_k}|V|^k)$. By summing over $k$ up to $s^*$, we obtain the total running time. Thanks to our heuristics (canonical forms, expansion strategy, candidate pruning, early termination), this time is significantly reduced in practice.

### A.3 Datasets

Table 2 reports the characteristics of the simplicial complexes used.

Table 2: Characteristics of the datasets used.

|  | DBLP | Enron | Zebra | ETFs | High | Primary |
| --- | --- | --- | --- | --- | --- | --- |
| Vertices | 1918581 | 50224 | 10112 | 2340 | 327 | 242 |
| Edges | 7683001 | 330179 | 110910 | 5543 | 5818 | 8317 |
| Triangles | 11350197 | 1431200 | 1737598 | 7130 | 2370 | 5139 |
| Maximal Simplices | 1730504 | 75296 | 8146 | 1125 | 4862 | 8010 |
| Max Dim | 17 | 64 | 62 | 19 | 5 | 5 |

**DBLP** (Benson et al. [9]) is a DBLP co-authorship simplicial complex: each simplex represents a publication and its vertices are the corresponding authors.

**Enron** (https://www.cs.cmu.edu/~enron) contains emails sent from employees of the Enron corporation. Vertices are senders and recipients, and simplices denote emails.

**Zebra** is constructed from genetic data of zebrafishes [38]. Starting from the gene correlation table (https://coxpresdb.jp/download), we obtain a simplex from each set of genes with mutual rank value lower than $\frac{1}{10}$th of the average value in its neighborhood. We retain the simplices with size in the 99th percentile.

**ETFs** (Exchange-Traded Funds - shorturl.at/itEK7) contains general aspects, portfolio indicators, returns, and financial ratios of 2353 ETFs, scraped from the Yahoo Finance website (https://finance.yahoo.com). Each simplex is a set of ETFs with Kendall correlation above the 99.9th percentile, where the correlation is computed exploiting the aforementioned features. We retain the simplices with size in the 99.9th percentile.

**High** [33] and **Primary** [21] are temporal simplicial complexes constructed from two SocioPattern datasets (http://sociopatterns.org). These datasets report active face-to-face contacts measured between students in a high school in Lycée Thiers, Marseilles, France; and children in a primary school in Lyon, France. Simplices are constructed by combining contacts that happen at the same time and location, and aggregate them into 10-minutes snapshots. The two complexes have 246 and 104 snapshots, respectively.